\documentclass{aastex}
\usepackage{spr-astr-addons}
\usepackage{url}

\RequirePackage{color}

\begin{document}

\title{A Possible Influence on Standard Model of Quasars and Active Galactic Nuclei in Strong Magnetic Field} \slugcomment{Not to appear in
Nonlearned J., 45.}
\shorttitle{A Possible Influence on Standard Model}
\shortauthors{Qiu-He Peng et al.}

\author{Qiu-He Peng \altaffilmark{1,3}, Jing-Jing, Liu \altaffilmark{2}, and Chi-Kang Chou \altaffilmark{4}}
\affil{Corresponding to Jing-Jing, Liu}
\email{qhpeng@nju.edu.cn, and liujingjing68@126.com}

\altaffiltext{1}{Department of Astronomy, Nanjing University,
Nanjing, Jiangshu 210000, China.}
\altaffiltext{2}{College of Marine
Science and Technology, Hainan Tropical Ocean University, Sanya,
Hainan 572022, China.}
\altaffiltext{3}{School of Physics and Space
Science, China West Normal University, Nanchong, 637009,China.}
\altaffiltext{4}{National Astronomical Observatory, Chinese Academy
of Sciences, Beijing, 100000, China.}

\begin{abstract}
Recent observational evidence indicates that the center of our Milky
Way galaxy harbors a super-massive object with ultra-strong radial
magnetic field \citep{b6}. Here we demonstrate that the radiations
observed in the vicinity of the Galactic Center (GC) \citep{b11}
cannot be emitted by the gas of the accretion disk since the
accreting plasma is prevented from approaching to the GC by the
abnormally strong radial magnetic field. These fields obstruct the
infalling accretion flow from the inner region of the disk and the
central massive black hole in the standard model. It is expected
that the observed radiations near the GC can not be generated by the
central black hole. We also demonstrate that the observed
ultra-strong radial magnetic field near the GC \citep{b6} can not be
generated by the generalized $\alpha$-turbulence type dynamo
mechanism since preliminary qualitative estimate in terms of this
mechanism gives a magnetic field strength six orders of magnitude
smaller than the observed field strength at $r=0.12$ pc. However,
both these difficulties or the dilemma of the standard model can be
overcome if the central black hole in the standard model is replaced
by a model of a  supper-massive star with magnetic monopoles (
SMSMM) \citep{b8}. Five predictions about the GC have been proposed
in the SMSMM model. Especially, three of them are quantitatively
consistent with the observations. They are: 1) Plenty of positrons
are produced, the production rate is $6\times10^{42} e^+
\rm{s}^{-1}$ or so, this prediction is confirmed by the observation
\citep{b33}; 2) The lower limit of the observed ultra-strong radial
magnetic field near the GC \citep{b6}, is just good agreement with
the predicted estimated radial magnetic field from the SMSMM model,
which really is an exclusive and a key prediction; 3) The observed
power peaking of the thermal radiation is essentially the same as
the theoretical prediction from the SMSMM model. Furthermore, the
observed ultra-strong radial magnetic field in the vicinity of the
GC may be considered as the astronomical evidence for the existence
of magnetic monopoles as predicted by the particle physics.
\end{abstract}

\keywords{quasars: general --- galaxies: magnetic fields ---
Physical Date and Processes: black hole physics.}


\section{Introduction}
It is now generally believed that bright quasars observed at large
red-shift are supermassive and rapidly spinning black holes formed
in the primordial universe. The spectacularly huge luminosity is
supplied by the black hole and the surrounding accretion disk. In
such models, magnetic fields play a very important role. More
specifically, the magnetic coupling between the central black hole
(or some supermassive stellar object) and the accretion disk enable
effective transport of energy and angular momentum between them. If
the black hole spins faster than the disk, energy and angular
momentum can be extracted from the black hole and transferred to the
disk vie the pointing flux \citep{b1,b2,b3}. It is now well
established that the transfer of energy and angular momentum in such
sping black hole and accretion disk system with magnetic coupling
can generate relativistic jet by the  mechanism of \citet{b2} if the
energy source is the spinning black hole.

The accretion model of GC is hot accretion flow. One of the most
important progress in this field is the finding of strong wind from
black hole hot accretion flow. For example, reformulated the
adiabatic inflow and outflow solution (ADIOS) model for radiatively
inefficient accretion flows, treating the inflow and outflow zones
on an equal footing. \citet{b37} presented the  results from two
long-duration general relativistic magneto-hydrodynamic (GRMHD)
simulations of advection-dominated accretion around a non-spinning
black hole. However, accretion onto a super-massive black hole of a
rotating inflow is a particularly difficult problem to study because
of the wide range of length scales involved. By using the ZEUS code,
\citet{b38} run hydrodynamical simulations of rotating, axisymmetric
accretion flows with Bremsstrahlung cooling, considering solutions
for which the centrifugal balance radius significantly exceeds the
Schwarzschild radius, with and without viscous angular momentum
transport.

On the other hand, Hydrodynamical (HD) and magnetohydrodynamics
(MHD) numerical simulations of hot accretion flows have indicated
that the inflow accretion rate decreases inward. Two models have
been proposed to explain this result. In the adiabatic
inflow-outflow solution (ADIOS), this is because of the loss of gas
in the outflow. In the alternative convection-dominated accretion
flow model, it is thought that the flow is convectively unstable and
gas is locked in convective eddies. Some authors have discuss
detaily these problems (e.g.,\citet{b39,b40,b41}). Based on the
no-outflow assumption, \citet{b42} investigated steady-state,
axisymmetric, optically thin accretion flows in spherical
coordinates. By comparing the vertically integrated advective
cooling rate with the viscous heating rate, They found that the
former is generally less than 30\% of the latter, which indicates
that the advective cooling itself cannot balance the viscous
heating.

Some researches show that the quasi-relativistic jets and
nonrelativistic wind may be produced by the joint action of
\citet{b1} and other mechanisms and the energy source in these cases
is rotational energy of the accretion flow originated from the inner
region of the disk \citep{b3}. Moreover, in magnetically arrested
disk models, the accreting plasma drags in a strong poloidal
magnetic field to the center such that the resulting accumulated
magnetic flux disrupts the axisymmetric accretion flow at a
relatively large magnetospheric radius \citep{b4}. Furthermore,
magnetic flux threading the black hole, rather than black hole spin
is the dominant factor in launching powerful jets and determining
the radio loudness of active galactic nuclei (AGN) \citep{b5}. Thus,
in the current models just briefly delineated above, magnetic fields
always play a key role. In the absence of these fields, it is almost
impossible to construct such useful, elegant and realistic models
for quasars and AGN.

More latest observational evidence indicates that the center of our
Milky Way harbours the closest candidate for a supermassive black
hole with strong magnetic fields. Using multi-frequency measurements
with several radio telescopes, \citet{b6} showed that there is a
dynamically relevant magnetic field near the black hole. If this
field is accreted down to the event horizon, it provides enough
magnetic flux to explain the observed emission from the black hole,
from radio to X-rays. In addition, \citet{b7} reported that jet
magnetic field and accretion disk luminosity are tightly correlated
over seven orders of magnitude for a sample of 76 radio-loud active
galaxies. They concluded that the jet-launching regions of these
radio-loud galaxies are threaded by dynamically important magnetic
fields, which will affect the disk properties. In this paper, we
investigate the plausible modifications of the standard models of
quasars and AGN in light of the very recent observational evidence
for the important discovery of a dynamically relevant magnetic field
near the GC. In particular, we focus on the possible origin of the
strong magnetic field in the galactic nucleus and study some of the
important effects of the ultra-strong magnetic field. The very
recent astronomical observations concerning the strongly magnetized
super-massive central black hole are depicted in considerable detail
in Section 2. The key roles played by such observed ultrastrong
radial magnetic fields in the standard models due to the effects of
these fields are elaborated in Section 3. The possible origin of
these strong magnetic fields near the GC will be considered in
detail in section 4. Our model of super-massive star with magnetic
monopoles (SMSMM)is given in Section 5. We show explicitly there the
generally accepted $\alpha$-turbulence dynamo mechanism of Parker
can not be used to generate the observed strong radial magnetic
field by a preliminary estimate in terms of the observed W51 data.
However, good agreement with observations may be achieved if the
central black hole of the standard model is replaced by a
supermassive stellar object containing magnetic monopoles
\citep{b8}. In these model, the production of the strong radial
magnetic fields can be naturally explained. We also discuss other
evidences against the black hole model of quasars and AGN in Section
6. Finally, in Section 7, we briefly summarize and emphasize our
results.


\section[]{Recent Astronomical Observations}

New progress and some discoveries of radio astronomical observations
near the GC have been reported in recent years. The main
performances of these are as follows:

(i) The measurement of an abnormally strong radial magnetic field
near the GC has been reported by \citet{b6}. The very important
results are as follows,  in particular, at $r=0.12$ pc, the lower
limit of the outward radial magnetic field near the GC is:
\begin{equation}
B\geq
8[\frac{RM}{66.960\rm{m^{-2}}}][\frac{n_e}{26\rm{cm^{-3}}}]^{-1}
\rm{mG},
 \label{1}
\end{equation}
where $n_e$ is the number density of electrons. At $r=0.12$ pc by
Chandra X-ray observation, $n_e\approx 26\rm{cm^{-3}}$\citep{b9}. A
theoretically calculated electron density of radioactively
inefficient accretion flows for the accretion disk around the GC) at
is about $20\sim100\rm{cm^{-3}}$, near the observational
value\citep{b10}.

It is well known that the interstellar magnetic field in the galaxy
is usually along the galactic spiral arms, and the average strength
of the magnetic field is about $1\rm{\mu}$Gauss. So the magnetic
field shown in Eq.(1) is abnormally strong.

(ii)Some radiation have been detected in the region near the central
region of the GC. We can see the review article written by
\citet{b11}. First, some radiations from the radio to the sub-mm
wavelength band have been detected in the region $(10-50) R_g$
around the central black hole with mass $4.3\times 10^6
\rm{M}_\odot$ for the GC. Second, Sgr A* is identified as a
surprisingly weak X-ray source by Chandra and it is inferred as
radiated from the region $\leq 10R_g$ due to the hour long timescale
for some detected weak X-ray flare and small NIR flare. Finally, The
radio flux density shows a flat-to-inverted spectrum, i.e., it rises
slowly with frequency with the power peaking around $10^{12}$Hz in
the sub mm band.

\begin{equation}
L_{\nu}=4\pi D^2_{\rm{Sgr A^*}}S_\nu,
 \label{2}
\end{equation}

At GHz frequencies, $S_\nu\propto \nu^{\alpha}, \alpha \sim 0.3\pm
0.1 $, The spectrum continues towards low frequencies (~300MHz) with
no sigh of absorption. At higher frequencies the spectrum extends
into Sub-mm wavelength regime, where the spectrum peaks and then
suddenly cuts-off \citep{b11}.

\section[]{The Effect of the Observed Strong Radial Magnetic Field on the Standard Model of Quasars and AGN}

The most important effect of the observed ultra-strong radial
magnetic field near the GC is that the assumption made in standard
models of the GC is invalid. This is because in the presence of the
strong radial magnetic field in the vicinity of the GC will prevent
material in the accretion disk from approaching to the GC due to the
magnetic freeze effect, when the kinematic energy density of the
material is less than the energy density of the magnetic field, or
the magnetic field is stronger than the Alphen critical value:

\begin{equation}
 B>B_{\rm{Alphen}}=(4\pi \rho v^2_{\rm{rot}})^{\frac{1}{2}},
 \label{3}
\end{equation}
where $v_{rot}$ is the rotation velocity of the accretion disk
around the GC and $\rho$ is the mass density. Making use of the
relation $v_{\rm{rot}}/c=\sqrt{(R_S/r)}$, we then have
\begin{eqnarray}
 B_{\rm{Alphen}}&=(4\pi \rho v^2_{\rm{rot}})^{\frac{1}{2}}=[\frac{4\pi nc^2}{N_A}]^{\frac{1}{2}}(\frac{R_S}{r})^{\frac{1}{2}}\nonumber\\
 &\sim
 13(\frac{n}{10^{4}\rm{cm^{-3}}})^{\frac{1}{2}}(\frac{R_S}{r})^{\frac{1}{2}}
 \rm{G}
 \label{4}
\end{eqnarray}

Here we note that magnetic field lines may be brought by a intense
turbulent plasma fluid.  Turbulent Richardson advection brings field
lines implosively together from distances far apart to separations
of the order of gyro radii \citep{b12}. This effect of advection in
rough velocity fields, which appear non-differentiable in space,
leads to line motions that are completely indeterministic or
'spontaneously stochastic'. The turbulent breakdown of standard flux
freezing at scales greater than the ion gyro radius can explain fast
reconnection of very large-scale flux structures, both observed
(solar flares and coronal mass ejections) and predicted (the inner
heliosheath, accretion disks, $\gamma$-ray bursts and so on). For
laminar plasma flows with smooth velocity fields or for low
turbulence intensity, stochastic flux freezing reduces to the usual
frozen-in condition \citep{b12}

No violent active phenomena such as explosion or mass ejection has
been detected in the region near $r\approx 0.12$ pc  , so the
foregoing usual frozen-in condition is valid. However, the lower
limit of the outward radial magnetic field at $r=0.12$ pc, from the
GC is $B\geq 8[RM/(66.960\rm{m^{-2}})][n_e/26\rm{cm^{-3}}]^{-1}
\rm{mG}$.

As is well known in the popular black hole model of the GC, the
radiation from the vicinity of the black hole originates from the
inflowing material of the accretion disk \citep{b3}. However, the
accretion plasma is clearly prevented from approaching to the GC by
the radial magnetic field in the region near $r\sim0.12$pc around
the GC at least as explicitly demonstrated above. Thus, the
accretion material can't reach the region near the central black
hole. Consequently, the radiations observed near the GC cannot be
emitted by the gas of the accretion disk. This is a dilemma of the
standard accretion disk model of black hole at the GC.

\section[]{The Origin of the Radial Magnetic Field and the $\alpha$-turbulence Dynamo Mechanism}

Now another important question is how to generate the strong
magnetic field near the GC by use of the known common conventional
physics.

A rotating magnetic instabilities (RMI) is proposed as a angular
momentum transfer mechanism in an ionized plasma stream for
producing a thin hot coronal plasma jet in the polar region out of
the accretion disk \citep{b13, b14}. The magnetic field will be
enhanced due to increases of RMI. Under normal circumstances, the
magnetic field may be enhanced to that the magnetic pressure reaches
at about $1/10$ of the hot gas pressure in general. But in a special
case, the magnetic pressure may reach at near the hot gas pressure.

The distribution of the enhanced magnetic field in the region
$100R_g$ ($R_g=R_S/2$ ) around the central black hole at the GC has
been shown in the Fig.4 of a review paper written by \citet{b3}. You
may see that the enhance of the magnetic field is in the polar thin
hot coronal plasma region perpendicular to the main accretion disk.
However, it is impossible to produce the observed strong radial
magnetic field ($\geq 8$mG ) at a distance $r=0.12$ pc from the GC
by RMI mechanism. The reasons are given as follows: 1) The plasma
with electron density, $n_e\approx26 \rm{cm^{-3}}$, is one on the
main accretion disk rather than one of the polar thin hot coronal
plasma outside the main disk. RMI mechanism is negligible for this
region; 2) The enhanced magnetic field is less than $1\mu$G , even
though RMI is taken into account.

The most famous dynamo known up to now is a type of
$\alpha$-turbulence dynamo mechanism firstly proposed by Parker in
1953 \citep{b15, b16} in the solar convection zone. The key idea of
the $\alpha$-turbulence dynamo mechanism is that the induced
electro-dynamic potential of turbulence is parallel to the magnetic
field \citep{b15}
\begin{equation}
\vec{\varepsilon}=\alpha\vec{B},
 \label{eq:5}
\end{equation}

\begin{equation}
\alpha\equiv \alpha(\sigma t_c, \overline{\vec{v}\cdot
\vec{\omega}})=-\frac{\sigma t_c}{3c}\overline{\vec{v}\cdot
\bigtriangledown \times \vec{v}}=-\frac{\sigma
t_c}{3c}\overline{\vec{v}\cdot \vec{w}},
 \label{eq:6}
\end{equation}
where $\vec{\omega}=\bigtriangledown \times \vec{v}$ is the curl of
the turbulent velocity of the fluid, and it is approximately
equivalent to the large-scale vortex rotational angular velocity.
$\sigma$ is the electrical conductivity of the fluid and $t_c$ is
the typical timescale of the turbulence. Eq.(5) shows that the
electro-dynamic potential of turbulence is proportional to the
scalar product of the turbulent velocity with vorticity.

In order to appreciate the important physical significance of the
¦Á-turbulence dynamo mechanism, it is appropriate at this point to
elaborate more clearly about the novel idea of the mechanism. It is
easy to visualize the production of the toroidal magnetic field by
the stretching poloidal field lines due to differential rotation, if
an astronomical body has some internal differential rotation and a
poloidal field that can be stretched. However, if the poloidal field
cannot be sustained, it will eventually decay and the production of
the toroidal field will also stop. In a famous paper, Parker argued
that in a rotating stellar object, turbulent convective motions
would be able to complete the cycle by generating a poloidal field
from a toroidal field. If the convection takes place in a rotating
stellar object, as a rising blob of plasma expands it feels a
Coriolis force so that the fluid motion become helical in nature.
The nearly frozen-in toroidal field is thus twisted so as to yield
poloidal field. The small scale poloidal loops so formed are
coalesced through reconnection to yield a large-scale poloidal
magnetic field because turbulent diffusion in partially ionized
plasma can smoothen out the magnetic fields of the loops. The
poloidal and toroidal field can sustain each other through a cyclic
feedback process. Thus, briefly, the poloidal field can be stretched
by the differential rotation to generate the toroidal field whereas
the helical turbulence associated with convection in a rotating
frame to give back a field in the poloidal plane.

Historically, the original treatment of Parker was very much based
on intuitive arguments. A formal and systematic approach was
developed later by Steenbeck, Krause and R\"{a}dler, known as mean
field magneto-hydrodynamics. The most important physical quantity in
this mathematical theory is the mean e.m.f. (or electrical
potential) $\vec{\varepsilon}$ induced by the fluctuating flow
$\vec{v}$ and magnetic fields $\vec{B}$, namely,
$\vec{\varepsilon}=\overline{\vec{v}\times \vec{B}}$, where the
overine denotes the ensemble average. For a homogeneous, weakly
anisotropic turbulence, the mean electromotive force in the
$\alpha$-turbulence dynamo mechanism is given by equations (5) and
(6). In particular, we note that the helical turbulent motions can
twist the toroidal field lines to produce the poloidal field as
mentioned before. It is the ¦Á-coefficient which encapsulates this
effect of helical motions in the mathematical theory.

Therefore, the principle of the $\alpha$-turbulent dynamo may be
briefly sketched as follows:  A toroidal magnetic field
$\Longrightarrow$ A toroidal electro-dynamic
potential$\Longrightarrow$ A toroidal current $\Longrightarrow$ A
poloidal magnetic field $\Longrightarrow$ A poloidal electro dynamic
potential $\Longrightarrow$ A poloidal current $\Longrightarrow$ A
toroidal magnetic field \citep{b17}.

In 1980s, the simulation showed that the intensity of the magnetic
flux tube on the surface of the sun is about $10^5$ G. Such a strong
magnetic tube should be formed at the bottom of the troposphere
rather than in the troposphere. The traditional $\alpha$-turbulence
generator cannot operate in such an intense circular magnetic field.
To give a more convincible explanation for the periodicity of the
sunspot, lots of similar $\alpha$ effect dynamo theories are
developed \citep{b18,b19,b20}. For example, the instability of
rotational magnetic field may magnify the magnetic field'. That
means the toroidal magnetic field is transformed to the poloidal
field due to the magnetofluid instability that results from the
interaction of the Coriolis force and the differential rotation (in
different latitude) in the sun. Generally speaking, the
magnification of the magnetic field is driven by the differential
rotation that interacts with the Coriolis force.

In fact, the Babcock-Leighton mechanism that is developed at the
same time as the  turbulent dynamo is recently regarded as the most
prospective explanation for the periodicity of the magnetic field
activity in the sun \citep{b21}. In accordance with the  turbulent
dynamo, the poloidal magnetic field is also generated by the
Coriolis force. But the Babcock-Leighton mechanism is the Coriolis
force acting on the magnetic flux tube in a large scale, leading to
the tilt angle that is observed on the surface of the sun in the
activation region. So the toroidal magnetic flux tube has a certain
poloidal component when it emerges on the surface of the sun. The
poloidal component is the result of the attenuation of the active
region. The authors in the paper \citep{b20} indicated that the
nonaxisymmetric instability of the toroidal magnetic flux tube in a
rotating star can provide a dynamo effect. This instability occurs
in the form of spiral waves. The increase in their amplitude causes
a phase shift between the disturbed magnetic field and the disturbed
flow field, which leads to the generation of an electric field in a
direction parallel to the undisturbed field. Coupled with the
differential rotation, this effect will produce a type of dynamo.
The difference between the traditional  turbulence dynamo and it is
the traditional turbulence dynamo cannot be applied to the magnetic
field that is quite strong while this new type of dynamo qualifies
in this case.

\citet{b19} propose an $\alpha \Omega$ flux-transport dynamo for the
Sun that is driven by a tachocline effect. This $\alpha$-effect
comes from the global hydrodynamic instability of latitudinal
differential rotation in the tachocline, as calculated using a
shallow-water model. Growing, unstable shallow-water modes
propagating longitudinally in the tachocline create vortices that
correlate with radial motion in the layer to produce a
longitude-averaged net kinetic helicity and, hence, an
$\alpha$-effect. It is shown that such a dynamo is equally
successful as a Babcock-Leighton-type flux-transport dynamo in
reproducing many large-scale solar cycle features\citep{b19}. The
success of both dynamo types depends on the inclusion of meridional
circulation of a sign and magnitude similar to that seen on the Sun.
Both $\alpha$-effects (the Babcock-Leighton-type and tachocline
$\alpha$-effect) are likely to exist in the Sun, but it is hard to
estimate their relative magnitudes. By extending the simulation to a
full spherical shell, It is shown that the flux-transport dynamo
driven by the tachocline  $\alpha$-effect selects a toroidal field
that is antisymmetric about the equator, while the Babcock-Leighton
flux-transport dynamo selects a symmetric toroidal field
\citep{b19}. Since our present Sun selects antisymmetric fields, the
tachocline $\alpha$-effect must be more important than the
Babcock-Leighton $\alpha$-effect.

These theories are still under further research and discussion. The
average electric potential generated by them (e.g., see the
discussions by \citet{b20}) is just proportional to the magnetic
field strength, and its direction is parallel to the magnetic field
direction. However, the proportional coefficient $\alpha$ in Eq.(5)
differs a lot in different theories \citep{b18}. We start the
discussion from Eq.(5). They are called by a joint name
$\alpha$-turbulent dynamo' while the proportional coefficient
$\alpha$ may have an uncertainty in 1-2 orders of magnitude. In the
theory of $\alpha$-turbulent dynamo, the energy of magnetic field
per unit volume is equal to the energy of induced electric current
per unit volume, thus

In the theory of $\alpha$-turbulent dynamo, the energy of magnetic
field per unit volume is equal to the energy of induced electric
current per unit volume, thus,

\begin{equation}
\frac{B^2}{4\pi}=ne\varepsilon=yne \alpha(\sigma t_c,
\overline{\vec{v}\cdot \vec{\omega}})B
 \label{eq.7}
\end{equation}
so we have
\begin{equation}
B=8\pi eyn\alpha(\sigma t_c, \overline{\vec{v}\cdot \vec{\omega}}),
 \label{eq.8}
\end{equation}
where $n$ is the number density of the plasma particles, $y$ is the
degree of ionization.

Some relevant data in the Sun are given as follows. The mass density
in the solar convection zone where dynamo mechanism is valid, is
$\rho\approx 8 \rm{g cm^{-3}}$, or the number density of particles
$n=5\times10^{24} \rm{cm^3}$; The maximum magnetic field in the
solar convection zone is $B_{\rm{max}}\sim 10^5$G (The details of
the differential rotation in solar interior can be found from the
following webpage of http$://$www.aip.de/en/press/images/

It is expected that the $\alpha$-turbulent dynamo model originally
developed by Parker valid in the solar convection region can be
applied also in accretion disks surrounding massive black holes. The
origin of the magnetic fields in the inner region of the accretion
disks may be explained in terms of the ¨»-turbulent dynamo model. To
estimate the value of ¨» in the inner regions near the GC we may
compare the value of  relevant parameters involved in the solar
convection zone with those in the star forming region in the
interstellar clouds. Using this method, it is plausible to believe
that the uncertainties of the turbulent velocity $\vec{v}$, he
electrical conductivity $\sigma$, the time scale for turbulence
$t_c$, and the vorticity $\vec{\omega}$ of the fluid in the solar
convection region (see \citet{b18}) would not seriously affect the
accuracy of our estimation for the magnetic field strength in the
inner region of accretion disk near the GC.

Assuming the validity of the $\alpha$-turbulence dynamo mechanism
similar to that in the solar convection zone, and comparing the
interstellar magnetic field strength  with that of the sunspot, we
may estimate the uncertainty of the value of $\alpha$ in terms of
the recent observation for the collapsing core W51e2 of the star
forming region \citep{b22}.

We deduce from Eq.(8)

\begin{equation}
 B=B_{\odot \rm{max}}\frac{n}{n_{\odot}}\frac{y}{y_{\odot}}\frac{\alpha(\sigma
t_c, \overline{\vec{v}\cdot \vec{\omega}})}{\alpha(\sigma t_c,
\overline{\vec{v}\cdot \vec{\omega}})_{\odot}}
 \label{eq.9}
\end{equation}

\begin{equation}
B\sim 10^{-19}\frac{n}{5\rm{cm^{-3}}}r(\vec{v}_{\rm{turb}}, \sigma
t_c) \rm{G},
 \label{eq.10}
\end{equation}
where $r(\vec{v}_{\rm{turb}}, \sigma
t_c)=\frac{n}{n_{\odot}}\frac{y}{y_{\odot}}\frac{\alpha(\sigma t_c,
\overline{\vec{v}\cdot \vec{\omega}})}{\alpha(\sigma t_c,
\overline{\vec{v}\cdot \vec{\omega}})_{\odot}}$.

Though the turbulent velocity in an interstellar cloud nay reach
$\vec{v}_{turb}\sim 10 \rm{km s^{-1}}$,  the curl of the turbulence
velocity in the interstellar cloud is far smaller than that in the
Sun, but both the typical timescale of turbulence, $t_c$ and the
electric conductivity ($\sigma$) may be much larger than those of
the sun. Thus, the value of the factor $\alpha(\sigma t_c,
\overline{\vec{v}\cdot \vec{\omega}})$ is rather uncertain.

However, we may estimate it as follows. In some interstellar cloud,
the strongest magnetic field may reach 20mG and the number density
is $n=2.7\times10^7 \rm{cm^{-3}}$ near the collapsing core W51e2.
Using Eq.(10), the uncertain factor, $r(\vec{v}_{turb}, \sigma
t_c)$, may be determined in terms of the observed data for W51e2
just delineated, thus
\begin{equation}
r(\vec{v}_{\rm{turb}}, \sigma t_c)\sim 3.7\times10^{10}
 \label{eq.11}
\end{equation}

In recent work by \citet{b23}, Submillimeter Array Observations of
magnetic fields in a $\rm{H}_2$ molecular cloud has been made. the
magnetic field is estimated about 1 mG, and the corresponding number
density is about $2.7\times10^5 \rm{cm^{-3}}$. Thus
\begin{equation}
r(\vec{v}_{\rm{turb}}, \sigma t_c)\sim 1.85\times10^{11}
 \label{eq.12}
\end{equation}

We can find that the difference of Eq.(11) with Eq.(12) is a factor
of 5 and it is in the uncertainty region of the $\alpha$-coefficient
(it is may be to reach to (1-2) orders of magnitude, see the
discussions from \citet{b16}).

Using Eq.(9) in eq.(11), with the observed electron number density
$n\sim26 {\rm{cm}}^{-3}$ at the distance of 0.12pc from GC
\citep{b6}, the resulting magnetic field is given by $B\leq 0.1¦Ì$G
which is five order of magnitude smaller than the observed lower
limit for the field strength, 8mG \citep{b6}. At this point, we
would like to mention that such strong magnetic field at the
distance of 0.12pc from GC can not be generated by the recent
Magnetically arrested accretion disk model\citep{b3}, although a
strong vertical bipolar magnetic field is pushed into the central
black hole by the thermal and ram pressure of the accreting gas and
the maximum magnetic field strength at the horizon
($R_S\approx10^{12}\rm{cm}$) is roughly $10^3$ G.

A broad class of astronomical accretion disks is shown to be
dynamically unstable to axisymmetric disturbances in the presence of
a weak magnetic field by \citet{b13,b14}, an insight with
consequently broad applicability to gaseous, differentially-rotating
systems. However, we have shown that the RMI \citep{b13} can not
generate the observed strong magnetic field. Also, the magnetic
field is not generated by $\alpha$-dynamo. In addition to the two
mechanisms of magnetic field amplification, \citet{b35} proposed
that the radial inward advection can also generate strong magnetic
field. \citet{b35} calculate the advection/diffusion of the
large-scale magnetic field threading an advection-dominated
accretion flow (ADAF) and find that the magnetic field can be
dragged inward by the accretion flow. The measurement of an
abnormally strong radial magnetic field near the GC has been
reported in 2013 by \citet{b6}. The lower limit of the outward
radial magnetic field near the GC is: B $\geq$ 8mG (at r=0.12 pc).
In our paper, we have discussed and demonstrated that the kinematic
energy density of the material is less than the energy density of
the magnetic field at r=0.12 pc.

On the other hand, the problems discussed and the results obtained
by \citet{b35} shall not apply to the content of our paper discussed
according to the astronomical observation. We know that for magnetic
field, the magnetic flux is conserved. For the radial magnetic
field, if the field is adverted inward by gas, Strength of B is
proportional to reciprocal of radius square. Therefore, the magnetic
field can be easily amplified by inward motion of gas. At this
point, we would like to mention that such strong magnetic field at
the distance of 0.12pc from the GC cannot be generated by the recent
Magnetically arrested accretion disk model \citep{b3} due to the
strength of B is proportional to reciprocal of radius square,
although a strong vertical bipolar magnetic field is pushed into the
central black hole by the thermal and ram pressure of the accreting
gas and the maximum magnetic field strength at the horizon
($R_S\approx10^{12}$ cm) is roughly $10^3$Gauss. Thus the inward
advection of magnetic field by gas from larger radii (0.12pc) can
produce magnetic field about 0.1¦ÌG much less than observed
$B\geq8$mG.

Even in strongly magnetized magnetically arrested disk (MAD) state,
the black hole can accrete gas through non-axisymmetric spiral
streams. Therefore, there should be radiation from the black hole
hot accretion flow. \citet{b51} investigated the dynamics and
structure of accretion disks, which accumulate a vertical magnetic
field in their centers by using two- and three-dimensional MHD
simulations. The central field can be built up to the equipartition
level, where it disrupts a nearly axisymmetric outer accretion disk
inside a magnetospheric radius, forming a magnetically arrested disk
(MAD). On the other hand, the black hole (BH) accretion flows and
jets are qualitatively affected by the presence of ordered magnetic
fields. \citet{b52} discuss fully 3D global general relativistic MHD
simulations of radially extended and thick (height H-to-cylindrical
radius R ratio of $|H/R|\sim 0.2-1$) accretion flows around BHs with
various dimensionless spins (a/M, with BH mass M) and with initially
toroidally dominated ($\varphi$-directed) and poloidally dominated
(R-z directed) magnetic fields. However, the above opinions
expressed have nothing to do with the problem of radial magnetic
field (i.e., $B\geq8$mG) at the distance of 0.12pc from the GC in
our paper.

In addition, someone once put forward whether the plasma or gas of
accretion disks can flow into a black hole along the radial magnetic
field at the center of the Milky Way galaxy. It is simple answer
that if there are the material flow along the radial magnetic field
to the center of the galaxy, the speed of the radial free fall
material will be close to the speed of light. Therefore, the strong
powerful radiation will be produced. But this kind of abnormal quite
powerful radiation has always been not observed. So this kind of
circumstance should be ruled out.

\section[]{Our Model of Super-massive Star with Magnetic Monopoles
(SMSMM)}

We note that the important discovery of very strong radial magnetic
field in the Vicinity of the GC is consistent with the prediction
from our model of SMSMM \citep{b24}. Thus, it is plausible to
believe that this is just the astronomical evidence needed for the
existence of  magnetic monopoles as predicted by the grand unified
theory of particle physics. In addition, the observed radiation from
radio to sub-mm wavelength band with power peaking around $10^{12}$
Hz in the sub mm band and the x-ray radiation near the GC are also
essentially in agreement with the prediction of our paper\citep{b8}.
In other words, the dilemma of the standard accretion disk model
with supermassive black holes at the GC would disappear.

We have investigated the model of SMSMM in a series of papers since
1985 \citep{b25,b26,b27,b28,b29,b8,b24, b30}, and the main ideas of
our model are as follows:

1) The fact that magnetic monopoles (M) may catalyze nucleons to
decay (the Rubakov-Callan (RC) effect, $pM\rightarrow e^+ \pi^0 M
(85\%)$  or $pM\rightarrow e^+ \mu^+ \mu^- M (15\%)$, with the
number of baryons being non-conserved) as predicted by the grand
unified theory of particle physics is invoked as the main energy
source of quasars and AGN. The supermassive central black hole in
the standard model is replaced by a supermassive object containing
magnetic monopoles. And the accretion disk acts only as a minor
energy supply.

2) The gravitational effect around the SMSMM in the galactic center
is similar to that around a massive black hole. However, the
supermassive object containing sufficient magnetic monopoles has
neither the horizon nor the central singularity. This is because the
reaction rate of the nucleon decay catalyzed by magnetic monopoles
is proportional to the square of mass density. Both the leptons and
photons from the decay are emitted outward, and the central density
cannot approach infinity. Combined with the RC effect from particle
physics, our model can avoid the central singularity problem in the
standard model of black hole theory.

On the other hand, some predictions about the GC in our model are as
follows\citep{b8}:

1) Plenty of positrons are generated and are emitted from the GC.
The producing rate is about $6.0\times10^{42}e^+\rm{s^{-1}}$. This
prediction is quantitatively confirmed by observation of high energy
astrophysics quantitatively ($3.4-6.30\times10^{42}e^+\rm{s^{-1}}$).

2) Some higher energy radiation above 0.5MeV may be emitted. The
integral energy of the high energy radiation is much higher than
both the total energy of the spectra of electron and positron
annihilation, and the total thermal luminosity of the central
object. This prediction is also consistent with observations.

3) The magnetic monopole condensed in the core region of the
supermassive object can generate radial magnetic field. The magnetic
field strength at the surface of the object is about 20-100 Gauss
(the radius of the object is about $8.1\times10^{15}$cm  or
$1.1\times10^4 R_S$ ($R_S$ is the Schwarzschild radius). We declared
previously in our article  that this prediction is the most crucial
one, which can be testified by future radio observations\citep{b8}.
Because the decrease of the magnetic field strength is proportional
to the inverse square of the distance from the source, so we have
$B\approx(10-50)$ mG at $r=0.12$ pc. This prediction is in agreement
with the lower limit of the observed  magnetic field (the detailed
discussions can be found from the article of \citet{b6}).

4) The super-massive objects containing saturated magnetic monopoles
in the centers of all the AGN in the region $D\leq50$ Mpc from the
earth may be the sources of observed ultra-high energy cosmic rays
($E_{\gamma}\sim (10^{18}-10^{21})$ eV), which can not be explained
up to date except our model \citep{b8}.

5) The surface temperature of the super-massive object in the
galactic center is about 120 K and the corresponding spectrum peak
of the thermal radiation is at $10^{12}$ Hz in the sub-mm wavelength
regime. This prediction is basically consistent with the recent
observation \citep{b11}.

The non- thermal radiation such as synchrotron radiation, may be
emitted due to the motion of the relativistic electrons in the
magnetic field. However, quantitative comparison of observations
with theory is rather difficult now, because the power indexes of
both the thermal radiation and the non- thermal radiation for the
radio wavelength band have not been well determined yet up to now.
The predictions 1), 3), 5) have been confirmed by the astronomical
or by the astrophysical observations in quantitative. Especially,
the third one is an exclusive prediction. It is hardly a
coincidence.

\section[]{Other Evidences Against the Black Hole Model of Quasars
and AGN}

We now briefly mention some other relevant evidences against the
black hole model of quasars and active galactic nuclei (AGN).

1) On strong radial magnetic fields in the galactic center for 76
radio-loud active galaxies

\citet{b7} did a statistical analysis on 76 radio-loud active
galaxies and concluded that there are very strong radial magnetic
fields in the galactic center preventing material in the accretion
disk from falling in, i.e. the accretion disk is not near the
central black holes in AGN. This could invalidate the standard
accretion disk model of black holes in AGN.

2) On two hot dust-free quasars

Using the Spitzer Space Telescope, \citet{b31, b32} discovered two
quasars without hot-dust emission in a sample of 21 ($z\approx 6$)
quasars by deep infrared photometry.

It is generally believed that quasars are powered by mass accretion
onto the central black holes and hot dust is directly heated by
quasar activity. So the discovery of the two hot-dust-free quasars
becomes a puzzle. The puzzle may be explained in terms of the
standard model for quasars with central black holes as follows.

We note that the two hot dust-free quasars with the lowest hot-dust
abundances have the smallest black hole masses
($(2-3)\times10^8\rm{M_{\odot}}$) and highest Eddington luminosity
ratios ($\sim2$) in the $z\approx 6$ sample, thus they are in an
early stage of quasar evolution with rapid mass accretion, but are
too young to have formed a detectable amount of hot dust around
them.

Since the accretion disk is not near the central black holes in the
quasars and AGN due to the presence of the observed strong radial
magnetic field near the GC, no material can reach the central region
of the quasars and AGN and the ultra -luminous radiation cannot be
emitted by the accretion disk model of black holes. The observations
of the two hot dust-free quasars may be considered as the important
evidence that the ultra -luminous radiation of the very young
quasars in the early stage of the universe cannot be emitted by the
accretion material flow from the accretion disk around, but it may
be produced by the RC effect in our AGN model containing magnetic
monopoles.

3) On the mass of the quasars

It is now generally believed by most astronomers that bright quasars
observed at large redshift (for example, $z >1$ or even $z > 5$) are
supermassive black holes ($m>10^{10} \rm{M_{\odot}}$) formed in the
primordial universe. The spectacularly huge luminosity is supplied
by the accretion of matter outside these black holes. As a result,
the mass of nearby galactic nuclei and quasars must be greater than
that of the remote quasars with larger redshift. This is because the
mass of the black holes must continuously increase due to accretion.
But the deduction is just contrary to the observation that no
supermassive black hole with mass $m>10^{9} \rm{M_{\odot}}$. Indeed,
this is the dilemma of the black hole model of quasars and active
galactic nuclei (AGN), although some proposals had been suggested
such as: (i) the merger of two galactic nuclei may also form a
larger quasar or an AGN; (ii) the mass of the supermassive black
holes at the center of AGN (and quasars) in the high red shift
region are much larger than those in the low red shift region is
only a select effect in observation because the supermassive quasars
are easy to observe due to their  huge luminosities. But, these
proposals cannot explain the fact that why no supermassive black
holes ($m>10^{9} \rm{M_{\odot}}$) have been found near the Milky Way
galaxy ( $D<1$ Gpc).

However, it is naturally explained by our AGN model containing
magnetic monopoles. The mass of the supermassive object must
decrease gradually due to the baryons decay catalyzed by the
magnetic monopoles and the decaying products (including  pions ,
muons , positrons and the radiation) are lost from the central
massive stellar object continuously. So the conclusion from our AGN
model containing magnetic monopoles is that the mass of the quasars
and AGN would decrease with the redshift (z). This is consistent
with observations.

4) On Jets of AGNs

\citet{b53}investigated the process of rapid star formation
quenching in a sample of 12 massive galaxies at intermediate
redshift ($z\sim 0.6$) that host high-velocity ionized gas outflows
($v>1000$ km s$^{-1}$).They conclude that these fast outflows are
most likely driven by feedback from star formation rather than
active galactic nuclei (AGN). They use multiwavelength survey and
targeted observations of the galaxies to assess their star
formation, AGN activity, and morphology. Common attributes include
diffuse tidal features indicative of recent mergers accompanied by
bright, unresolved cores with effective radii less than a few
hundred parsecs. The galaxies are extraordinarily compact for their
stellar mass, even when compared with galaxies at $z \sim 2-3$. For
$9/12$ galaxies, we rule out an AGN contribution to the nuclear
light and hypothesize that the unresolved core comes from a compact
central starburst triggered by the dissipative collapse of very
gas-rich progenitor merging disks. They find evidence of  AGN
activity in half the sample but we argue that it accounts for only a
small fraction ($\sim 10\%$) of the total bolometric luminosity.
They find no correlation between AGN activity and outflow velocity
and they conclude that the fast outflows in our galaxies are not
powered by on-going AGN activity, but rather by recent, extremely
compact star-bursts.

Numerical simulations of hot accretion flows around black holes have
shown the existence of strong wind \citet{b54, b55}. They performed
hydrodynamic (HD) simulations and the MHD simulations by taking into
account the gravitational potential of both the black hole and the
nuclear star cluster.  they found that, just as for the accretion
flow at small radii, the mass inflow rate decreases inward, and the
flow is convectively unstable. However, a trajectory analysis shows
that there is very little wind launched from the flow. They
concluded the result that wind cannot be produced in the region $R>
R_A$ (here $R_A$ is similar to the Bondi radius). The winds are
mainly determined by the galactic nuclei clusters of gravitational
potential. Near the event horizon surface accretion flow is
determined by the black hole's gravitational potential. Their
results shown that Jets of active galactic nuclei (i.e., Jet,
stellar winds) is actually dominated by the gravitational potential
of the galactic nuclei clusters, and has nothing to do with the
central black hole. This is very obvious that they do not support
the popular view, which the active galactic nuclei jets are strong
evidence of black holes model.

\section{Conclusions}

In conclusion, we have demonstrated that the radiations observed in
the contiguous region of the central black hole cannot be emitted by
the gas of the disk since the accreting plasma is prevented from
approaching to the GC by the ultrastrong magnetic fields. In
addition, we have also shown that the observed strong radial
magnetic fields near the GC by \citet{b6}, cannot be generated by
the ¨»-turbulence dynamo mechanism of Parker because qualitative
estimate gives a magnetic field strength six orders of magnitude
smaller than the observed field strength at $r=0.12$ pc. The dilemma
of the standard model for quasars and AGN can be avoided if the
central black hole in the standard model is replaced by a
supermassive stellar object containing magnetic monopoles. The
radiations emitted from the inner region of the galactic nucleus and
the discovery of the strong radial magnetic field near the GC can
all be naturally explained by our model\citep{b8}. Moreover, the
observed ultra-strong radial magnetic field in the vicinity of the
GC may be considered as important astronomical evidence for the
existence of magnetic monopoles as predicted by the Grand Unified
Theory of particle physics.

\acknowledgments We would like to thank Prof. Daniel Wang, Prof.
Y.F. Huang, Prof. P.F. Chen and Prof. J. L. Han for their help to
inform us some new information of observations. This work was
supported in part by the National Natural Science Foundation of
China under grants 10773005, 11273020, 11565020, and the Counterpart
Foundation of Sanya under grant 2016PT43, the Special Foundation of
Science and Technology Cooperation for Advanced Academy and Regional
of Sanya under grant 2016YD28, and the Natural Science Foundation of
Hainan province under grant 114012.

\end{document}